\documentstyle[12pt]{article}
\input epsf.tex
\begin{document}

%  Greek letters
\def\a{\alpha}
\def\b{\beta}
\def\ch{\chi}
\def\d{\delta}
\def\e{\epsilon}
\def\f{\phi}
\def\g{\gamma}
\def\h{\eta}
\def\i{\iota}
\def\j{\psi}
\def\k{\kappa}
\def\l{\lambda}
\def\m{\mu}
\def\n{\nu}
\def\o{\omega}
\def\p{\pi}
\def\q{\theta}
\def\r{\rho}
\def\s{\sigma}
\def\t{\tau}
\def\u{\upsilon}
\def\x{\xi}
\def\z{\zeta}
\def\D{\Delta}
\def\F{\Phi}
\def\G{\Gamma}
\def\J{\Psi}
\def\L{\Lambda}
\def\O{\Omega}
\def\P{\Pi}
\def\S{\Sigma}
\def\U{\Upsilon}
\def\X{\Xi} 
\def\T{\Theta}
\def\vf{\varphi}

\def\Ab{\bar{A}}
\def\gi{g^{-1}}
\def\li{{ 1 \over \l } }
\def\lb{\l^{*}}
\def\zb{\bar{z}}
\def\ub{u^{*}}
\def\vb{v^{*}}
\def\Tb{\bar{T}}
\def\pp {\partial }
\def\pb {\bar{\partial }}
\def\be{\begin{equation}}
\def\ee{\end{equation}}
\def\ben{\begin{eqnarray}}
\def\een{\end{eqnarray}} 
\def\tr{{\rm{tr}}} 
\def\ddt{\hat}
\def\dddt{\acute}

\hsize=16.5truecm
\addtolength{\topmargin}{-0.8in}
\addtolength{\textheight}{1in}
%\vsize=26truecm
\hoffset=-.5in

\thispagestyle{empty}
\begin{flushright} \ December \ 1996\\
SNUCTP 97-080\\
\end{flushright}
\begin{center}
 {\large\bf Composite Skyrme Model with Vector Mesons  }\\[.1in]
\vglue .5in
 Kwanghoon Hahm, \ \ KyoungHo Han\footnote{ E-mail address; khan@photon.kyunghee.ac.kr }
\\[.2in]
{and}
\\[.2in]
H. J. Shin\footnote{ E-mail address; hjshin@nms.kyunghee.ac.kr }
\\[.2in]
{\it  
Department of Physics \\
and \\
Research Institute of Basic Sciences \\
Kyunghee University\\
Seoul, 130-701, Korea}
\\[.2in]
{\bf ABSTRACT}\\[.2in]
\end{center}
We study the composite Skyrme model, proposed by Cheung and G\"{u}rsey, introducing
vector mesons in a chiral Lagrangian. We calculate the static properties of
baryons and compare with results obtained from models without vector mesons. 
\vglue .1in

\newpage
More than thirty years ago, Skyrme\cite{skyrme} pointed out that the
non-linear sigma model\cite{sigma} 
with a quartic derivative term admits stable soliton solutions which he
suggested to identify as the baryons. Starting from 
this consideration, the phenomenological 
aspects of the Skyrme soliton have been widely explored in the past years.
It was shown the computed static properties of 
the $SU(2) \times SU(2)$ Skyrme model agree
moderately well with experiments.\cite{adkins} An example was
the predicted pion decay constant which is 30\% lower than the 
experimental value. Various attempts such as adding 
higher-order terms to the Lagrangian\cite{omega}, 
including the effect of the pion mass\cite{adkins5}, 
and extending the symmetry to $SU(3) \times SU(3)$\cite{pras} 
have failed to increase
significantly the value of the pion decay constant.

The work of Cheung and G\"{u}rsey generalizes the Skyrme model by introducing $U$ 
fields, for which $U^n$ not $U$ transforms linearly under $SU(2) \times SU(2)$.\cite{gursey} 
%%%%%%% 
Due to this invariance of the generalized model is not modified
the current algebra result with the new parameter n.
%%%%%%%
They found this modification without any tricks gives for n=3 almost 
experimentally observed value for the pion decay constant $F_\p$ to be
185$MeV$. The fact that n=3 is singled out led these authors to consider
the possibility that this composite Skyrme model describes effective quarks
in a nucleon. Moreover some other predicted static properties were shown
to agree better
with the experiment than the original model(n=1), while a disagreement persists
for the magnetic properties.

After the work of Ref.\cite{gursey}, Nam and Workman study the effects of
chiral symmetry breaking in the composite model.\cite{nam} But the predictions for
the magnetic properties were not improved noticeably and 
the fit to the pion decay constant even got worse. 

In this paper we introduce the $\o$ vector mesons in the composite Skyrme model
and calculate the static properties of a nucleon including the maganetic
properties. This problem is interesting in light of the fact that 
the incorporation of additional low-lying mesons into the effective
Lagrangian is the most important modification to construct a realistic theory
and make predictions more accurate. In fact the static properties of the nucleon
in the ``$\o$-stabilized" Skyrme
model(n=1) of Adkins and Nappi\cite{omega} constituted an improvement over
the unadulterated Skyrme model. Especially the prediction for the 
magnetic moment of the
proton(neutron) was improved from 1.97(-1.24) to 2.34(-1.46) while the
experimental value is 2.79(-1.91). There has been much work to extend the
nonlinear $\s$ model to the low-lying vector meson resonances such
as the $\o, \r$, and $A_1$.\cite{gen}\cite{gen1} In this respect the models
in this paper are certainly incomplete insofar as they include the
$\o$ meson only but this clearly has the advantage that one is able
to obtain in a simpler way guiding results for a full-scale calculation.

The object of this paper is then to have a global view of the 
``$\o$-stabilized" composite Skyrme model, with and without the pion mass term.
The Lagrangian for our purpose is described in terms of composite chiral
field $U^n$ coupled to an $\o$ vector meson field:
\ben
{\cal L} = -{1 \over 4} (\pp_\m \o_\n -\pp_\n \o_\m)(\pp^\m \o^\n-\pp^\n \o^\m)
+{1 \over 2} m_\o^2 \o_\m \o^\m + \b \o_\m B^\m & & \nonumber \\
+{1 \over 16n^2} F_\p^2 \tr(\pp_\m U^n \pp^\m U^{-n}) +{1 \over 8n^2}
F_\p^2 m_\p^2(\tr U^n-2),& & 
\een
where n is an arbitrary positive integer.
Here $m_\o$ is the mass of the omega, $m_\p$ the mass of the pion and
$B^\m$ the normalized baryonic current
\be
B^\m = {1 \over 24\p^2 n} \e^{\m\n\a\b} \tr[(U^{-n} \pp_\n U^n)
(U^{-n} \pp_\a U^n)(U^{-n} \pp_\b U^n)].
\ee
The model still retains the simplicity of the effective theory and the
original model of Adkins and Nappi\cite{omega} is contained as a 
special case of n=1.
%%%%%%%%%%
Note that the choice of the coefficients in the Lagrangian (1) guarantees that 
the composite model has the same chiral Lagrangian description of the pion-pion 
interaction, etc. as in conventional model. In fact when we expand the 
kinematic part of the Lagrangian, we can obtain the usual kinematic term
using
\be
U= \exp\left(2 i {\vec\t} \cdot {\vec\f}/F_{\pi}\right)
 \approx 1+ 2 i {\vec\t} \cdot {\vec\f}/F_{\pi} \ .
\ee
Similarly when we consider the coupling of the pions to the W gauged bosons,
the coeffcient of the term $\pp_{\m}{\vec\f}\cdot W^\m$ in 
$ (\pp_{\m} + i g {\vec\t} \cdot {\bf W_\m})U^n$
which gives the transition amplitude for the pions on the W's remains not 
modified with the paprameter n.
%%%%%%%%%%
The constant $\b$ is related to the strength of the coupling of the
omega to three pions, but we will take it as a free parameter and
determine it by fitting the properties of the nucleons following
the treatment of Ref.\cite{omega}.

As usual, we make the hedgehog ansatz for the soliton solution
\be
U({\bf x}) =U_0({\bf x}) = \exp[i {\bf \t \cdot \hat{r}} F(r)],
\ee
with the boundary condition $F(0)=\p, F(\infty)=0$. When we rotate
the soliton with a time dependent $SU(2)$ matrix $A(t)$:
\be
U(t,{\bf x})=A(t) U_0({\bf x}) A^{-1}(t),
\ee
we obtain the Lagrangian
\be
L=-M + I \tr[\pp_0 A \pp_0 A^\dagger]
=-(4\p F_\p ^2 /m_\o) \tilde{M} + (2 \p F_\p ^2/3m_\o^3) \tilde{I}
\tr[\pp_0 A \pp_0 A^\dagger].
\ee
Here $\tilde{M}$ and $\tilde{I}$ are the dimensionless numbers related
to $F(r)$ and $\o = F_\p \tilde{\o}$ as
\ben
\tilde{M} &=& \int_0 ^\infty dt [ -{1 \over 2} t^2 \tilde{\o}'^2
-{1 \over 2} t^2 \tilde{\o}^2 + \bar{\b} \tilde{\o} (\sin ^2 nF) F'
+ {1 \over 8}t^2 F'^2 +{1 \over 4n^2} \sin^2 nF \nonumber \\
 & &+{1 \over 4n^2} (m_\p ^2 /m_\o ^2) t^2 (1-\cos nF)] \nonumber \\
\tilde{I} &=& {1 \over n^2} \int_0 ^\infty t^2 \sin^2 nF dt 
+2 \bar{\b}^2 \int_0 ^\infty dt dt'
[\sin^2 nF(t)] F'(t) [\sin^2 nF(t')]F'(t') B(t,t'),
\label{dnum}
\een
where 
\be
B(t,t') =\{ \exp[-(t+t')](1+t+t'+tt')- \exp(-|t-t'|)(1+|t-t'|-tt')\}/tt'
\ee
and $\bar{\b} =\b m_\o /2 \p^2 F_\p$, $r=t/m_\o$. 

From $\tilde{M}$ in eq.(7) we
can numerically find $F(t)$ and $\tilde{\o} (t)$ by employing
relaxation techniques described in Ref.\cite{numer}. In Figs.1 and 2 we plot the solutions
$F(t)$ and $\tilde{\o} (t)$ for the models with the pion mass term.
In these figures $\bar{\b}$ is chosen to fit the experimental value of the mass ratio
\be
\tilde{M} / \tilde{I} =(5m_N- m_\D)(m_\D - m_N)/36 m_\o ^2 =0.04605,
\ee
which is from the formula for the energy $E= M + J^2/2I$. Some obtained 
$\bar \b$ values
are 4.9, 7.5, 8.6 for n=1, 4, 7. The pion decay constant $F_\p$ corresponding
to these values are 124, 121 and 120 MeV, showing essentially no dependence on n.
When we apply this procedure on models without the pion mass term we obtain
$\bar \b$=3.3, 2.0, 1.5 for n=1, 4, 7 with the pion decay constant 145, 194 and 219 MeV.

In order to compute some integrals for physical quantities with 
arbitrary values of n, we use following formulae:
\ben
\int d \O V^{0,a} &=& {4 \over 3} \p i [ F_\p^2 {\sin^2 nF(r) \over n^2}
 + 2 ({\b \over 2 \p^2})^2 {\sin^2nF(r) \over r^2} F'(r) \nonumber \\
 && \times \int_0^\infty dr' \sin^2 nF(r') F'(r') 
{B(m_\o r,m_\o r')\over m_\o} ] 
\tr[(\pp_0 A)
A^\dagger \t^a]
\een
\be
\int d\O {\bf q \cdot x} V^{i,a}  = {2\p \over 3} ({1 \over 2n^2} F_\p ^2
+{\b \o\over \p^2} F')(\sin^2 nF) \tr [{\bf \t \cdot q} \t^i A^\dagger \t^a
A],
\ee
\be
\int d\O A^{i,a} = {4 \over 3} \p [{1 \over 4} F_\p ^2({F'\over n} +{\sin 2nF \over n^2r
}) +{\b \o\over 2 \p^2} ({\sin 2nF \over r} F' +{\sin^2 nF \over nr^2})]
\tr [\t^i A^\dagger \t^a A].
\label{aint}
\ee

\begin{figure}
\centerline{\epsfxsize 4.5 truein \epsfbox {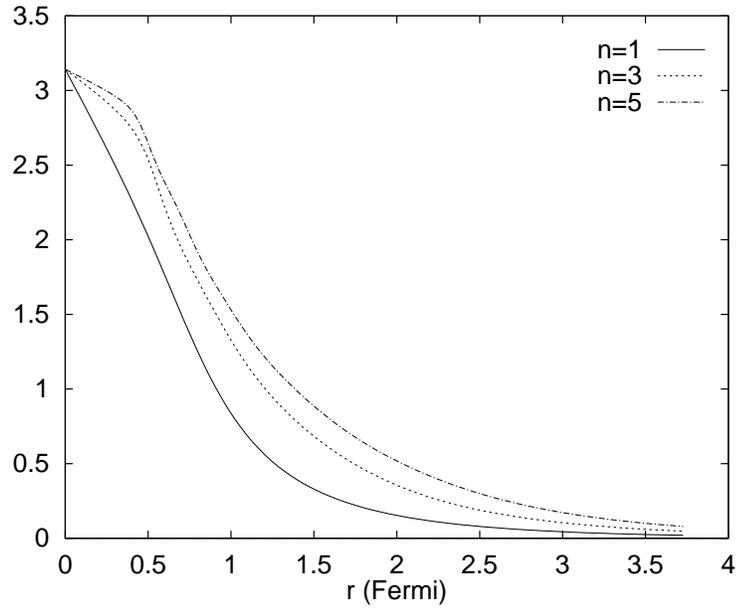}}
\caption{The shape function $F(r)$ for values of n=1 $\bar{\b}$=4.9, n=3
$\bar{\b}$=6.9, n=5 $\bar{\b}$=8.0.}
\end{figure}
\begin{figure}
\centerline{\epsfxsize 4.5 truein \epsfbox {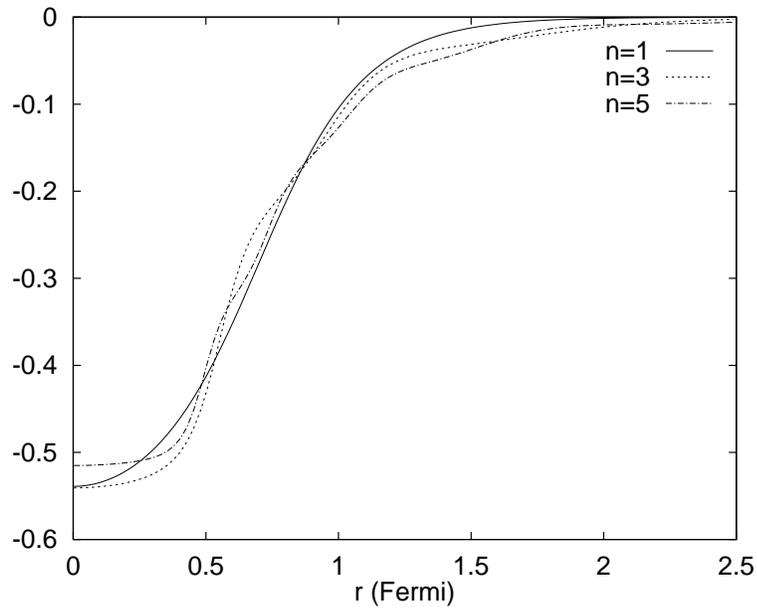}}                          
\caption{The omega field $\tilde{\o} (r)$ for values of n=1 
$\bar{\b}$=4.9, n=3          
$\bar{\b}$=6.9, n=5 $\bar{\b}$=8.0.}
\end{figure}

We evaluate various static properties of nucleons using these formulae and
compare them with the previous predictions of the Skyrme models. 
Table 1 shows the results from the broken chiral composite models having vector mesons 
and the pion mass term(BCVM) for n=1, 2 and 3.
Note that n=1 case corresponds to the calculation of 
Adkins and Nappi.\cite{omega}
They are compared with the predictions of the original(n=1$^*$) and
composite(n=3$^*$) Skyrme model having the Skyrme term and the
pion mass term(BCST).
(results of n=1$^*$ are from \cite{adkins5} and results of n=3$^*$ are from
\cite{nam}.) The fourth row shows $\bar \b$ values for BCVM models, 
while it shows the $e$ values in the coefficient
of Skyrme term for BCST models.\cite{adkins5}\cite{nam}
Similarly Table 2 shows results from the chiral theories
having no pion mass term. Here results from the models with vector mesons(CVM) 
for n=1, 2, 3 are displayed in the first three
columns  while results from the original model 
(results of n=1$^*$  are from \cite{adkins}.)
and composite model(results of n=3$^*$  are from 
\cite{gursey}\cite{nam}\cite{lee}.) having the Skyrme term(CST) are displayed in
the next two columns.
\begin{table}[b]
\caption{Results for broken chiral theories with(BCVM) and without vector mesons(BCST).}
\begin{tabular}{ccccccc} \hline\hline
Quantity&n=1&n=2&  n=3  &n=1$^*$&n=3$^*$& Experiment \\ \hline
  $M_N$                            & Input & Input & Input &  Input&Input  & 938.9 MeV \\
  $M_\D$                           &Input  & Input & Input & Input &Input  & 1232 MeV \\
  $F_\p$                           &124 MeV&124 MeV&123 MeV&108 MeV&127 MeV& 186 MeV \\
  $\bar \b$ or e                   &4.9    &6.1    &6.9    &4.84   &1.30   &          \\
  $\left<r^2\right>_{E,I=0} ^{1/2}$&0.74 fm&1.01 fm&1.19 fm&0.68 fm&0.99 fm&0.72 fm\\
  $\left<r^2\right>_{E,I=1} ^{1/2}$&1.06 fm&1.43 fm&1.65 fm&1.04 fm&1.65 fm&0.88 fm\\ 
  $\left<r^2\right>_{M,I=0} ^{1/2}$&0.94 fm&1.41 fm&1.76 fm&0.95 fm&1.29 fm&0.81 fm\\ 
  $\left<r^2\right>_{M,I=1} ^{1/2}$&1.03 fm&1.38 fm&1.58 fm&1.04 fm&1.28 fm&0.80 fm\\
  $\m_p$                           &2.33   &2.71   &3.03   &1.97   &2.37   &2.79   \\ 
  $\m_n$                           &-1.46  &-1.10 &-0.80   &-1.24  &-0.84  &-1.91  \\ 
  $g_A$                            &0.83   &1.28   &1.64   &0.65   &1.4    &1.23   \\ 
  $g_{\p NN}$                      &12.5   &19.4  &24.9  &11.9   &21     &13.5   \\ 
  $\s$                             &54 Mev &61 MeV &59 MeV &49 MeV &46 MeV&36$\pm$20 MeV\\ 
  $g_{\p N \D}$                    &18.8   &29.2   &37.4   &17.8   &31.5   &20.3   \\ 
  $\m_{N \D }$                     &2.67   &2.70   &2.71   &2.3    &2.27   &3.29   \\ 
  \hline\hline
\end{tabular}
n=1$^*$ and n=3$^*$ denote the results for broken chiral theories 
without vector mesons.
\end{table}

As observed in previous studies, the pion mass term perturbs the
chiral Skyrme soliton such that the value of $F_\p$ is less than that found in
chiral models. The excellent agreement of predicted $F_\p$ with the 
experimental value in the chiral composite model(CST) is still seen in
CVM model when n=3, while the predictions from 
broken chiral theories(BCVM or BCST)
lie below the experimental value. They show small dependence on n.
The various mean radii increase with n in all models and the addition of pion mass
term increase them further. The models with vector mesons(CVM or BCVM)
show steeper increase of mean radii with n than the models without
vector mesons(CST or BCST).
Generally the predictions for mean radii lie above the experimental values
in all models.

The predictions for the proton magnetic moment increase with n and approach the
experimental value in 
all models. Especially the results from CVM and BCVM are substantial
improvements on those from CST and BCST. But the results from CVM,
which are 2.18, 2.37, 2.45, 2.56 for n=1, 4, 7, 10,
still lie below the experimental value 2.79.
The best fit is obtained when n=2
for BCVM model with the calculated value 2.71. 
In the case of the neutron,  the predicted values decrease with n 
and the best result -1.53 is
obtained with CVM model when n=1. Thus the addition of vector mesons to n=1 models
improves the result substantially both for the proton and neutron magnetic moment, while
that which improves the result for $\m_p$ degrades the result for $\m_n$ for CVM and BCVM models
when n$>$1. This fact is related with the results for $\m_{N \D}$.
The calculations are 2.5-2.7
in CVM or BCVM models and 2.3 in CST or BCST models, showing small dependence on n.
Since $\m_{N \D}$ is related to $\m_p -\m_n$, the improvement of $\m_p$ with n
results in the deterioration of $\m_n$.
\begin{table}[b]
\caption{Results for chiral theories with(CVM) and without vector mesons(CST).}
\begin{tabular}{ccccccc} \hline\hline
Quantity&n=1&n=2&  n=3  &n=1$^*$&n=3$^*$& Experiment \\ \hline
  $M_N$                            & Input & Input & Input &  Input&Input  & 938.9 MeV \\
  $M_\D$                           &Input  & Input & Input & Input &Input  & 1232 MeV \\
  $F_\p$                           &145 MeV&166 MeV&182 MeV&129 MeV&185 MeV& 186 MeV \\
  $\bar \b$ or e                   &3.3    &2.8    &2.3    &5.44   &1.72   &          \\
  $\left<r^2\right>_{E,I=0} ^{1/2}$&0.64 fm&0.79 fm&0.85 fm&0.59 fm&0.74 fm&0.72 fm\\
  $\left<r^2\right>_{M,I=0} ^{1/2}$&0.91 fm&1.32 fm&1.61 fm&0.92 fm&1.18 fm&0.81 fm\\ 
  $\m_p$                           &2.18   &2.31   &2.32   &1.87   &2.03   &2.79   \\ 
  $\m_n$                           &-1.53  &-1.33  &-1.17  &-1.31  &-1.17  &-1.91  \\ 
  $g_A$                            &0.78   &1.25   &1.60   &0.61   &1.30   &1.23   \\ 
  $g_{\p NN}$                      &10.1   &14.1   &16.4   &8.9    &13.2   &13.5   \\ 
  $g_{\p N \D}$                    &15.2   &21.1   &24.7   &13.2   &19.8   &20.3   \\ 
  $\m_{N \D }$                     &2.62   &2.57   &2.47   &2.3    &2.26   &3.29   \\ 
  \hline\hline
\end{tabular}
n=1$^*$ and n=3$^*$ denote the results for chiral theories without vector mesons.
\end{table} 
The axial coupling $g_A$, which is calculated 
by integrating the axial current (12), increases with n
for all models. The best
result is given when n=2 for CVM or BCVM models and when n=3
for CST or BCST models. Another quantity which is
related to $g_A$ by  the Goldberger-Treiman relation  is
the pion nucleon coupling constant $g_{\p NN}$. For theories with massless pions
we calculated them from the long-distance behaviour of the shape function
$F(r)$\cite{adkins}, with the results in Table 2. Li and Liu argued that they can not be determined by the asymptotic
behaviour of $F(r)$ for theories with massive
pions.\cite{li} Instead we use the formula
\be
g_{\p NN} = {4 \p \over 9n} m_\p ^2 m_N F_\p \int r^3 \sin n F(r) dr,
\ee
with the results shown in Table 1.  It is explicitly
checked in both cases that they satisfy the Goldberger-Treiman relation 
within the numerical errors.
The calculated values of $\s$ slightly decrease with n in BCVM such as 
54, 61, 59, 56, 53, 50, 47, 45 MeV for n=1, 8, and approach
the experimental value 36$\pm$20 MeV. The results are reasonable
 but are slightly larger than those from BCST.  

The overall results from CVM or BCVM models seem to be more or less
better than those  from CST or BCST models. 
Especially the result for $\m_p$
is improved quite a lot while $\m_n$ is not. It was already observed
that when we are away from the chiral
limit, the worse numerical behaviour is resulted in the composite model.\cite{nam}
 The addition of
vector mesons in the composite models does not change this behaviour
and the best fit from the composite
models having vector mesons is obtained
 with the chiral theory(CVM) when n=3. Another 
strange prediction of the composite models was 
the peculiar oscillating behaviour of the nucleon charge density
which is not seen experimentally.
This behaviour still persists in CVM and BCVM models which we show in Fig. 3. 
\newpage
Generally the
composite Skyrme models have better description of nucleon properties
compared to the original n=1 model, 
but it still needs more improvements
to increase the phenomenological power.
One possibility is to use different integers $n$ in $U^n$ for
each term in the effective Lagrangian $\cal{L} = {\cal{L}}_{\rm 0} +
\cal{L}_\r + \cal{L}_\s + \cal{L}_\o$ of Ref.\cite{gen1}.

\begin{figure}
\centerline{\epsfxsize 4.5 truein \epsfbox {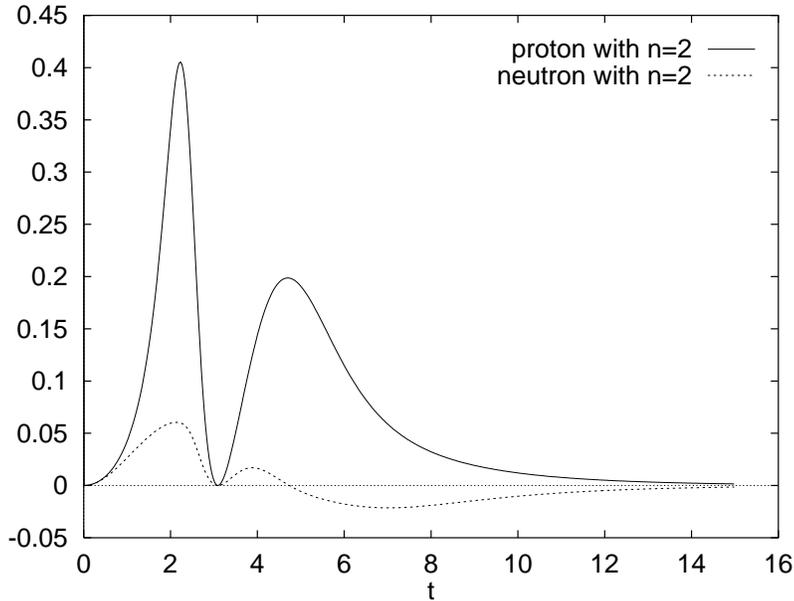}}
\caption{The charge density $\r (t)$ for the proton and the neutron
plotted as a function of the dimensionless variable $t$ with n=2
$\bar \b$=6.1. The density
functions are normalized to give a unit charge for the proton.}
\end{figure}

\vglue .3in 
{\bf ACKNOWLEDGEMENT}
\vglue .2in
We would like to thank to S. Nam and  B. Lee for many useful discussions. 
This work was supported in part by the program of Basic Science Research, 
Ministry of Education BSRI-96-2442, and by Korea Science and Engineering 
Foundation through CTP/SNU.

\vglue .2in

\end{document}